# A Basic Unified Context for Evaluating the Beam Forming and MIMO Options in a Wireless Link

Dmitry Chizhik, *Senior Member, IEEE*, Gerard J Foschini, *Life Fellow, IEEE* and R. A. Valenzuela, *Fellow, IEEE*
Bell Laboratories, Alcatel-Lucent, Holmdel, New Jersey, USA

*Abstract*- For one isolated wireless link we take a unified look at simple beamforming (BF) as contrasted with MIMO to see how both emerge and under which conditions advantage goes to one or the other. Communication is from a high base array to a user in clutter. The channel propagation model is derived from fundamentals. The base knows the power angular spectrum, but not the channel instantiation. Eigenstates of the field spatial autocorrelation are the preferred apodizations (APODs) which are drivers of the natural modes for exciting lectric fields. Preference for MIMO or BF depends on APOD spectra which are surveyed pointing to various asymptotic effects, including the maximum BF gain.
Performance is studied under varying eigenmode power settings at 10% outage. We focus on (1,4) driving the strongest mode for BF and (4,4) driving the 4 strongest for MIMO. Results are obtained under representative parameter settings, eg, an angular spread of 8º, 2 GHz carrier, 0 dB SNR and an array aperture of 1.68m (4 field decorrelation lengths) with antenna elements spaced as close as $\lambda/2$. We find MIMO excelling for array apertures much larger than the decorrelation length; BF does almost as well for smaller apertures.

## I. INTRODUCTION

### A. Motivation, Scope and Overview

We are interested here in evaluating the efficacy of beamforming (BF) and MIMO in a downlink environment for communication from a base to a user in clutter. Our setup includes a mathematical model of linear antenna arrays and the key statistical properties of the correlated field along the transmit array aperture. We will explore how to power these arrays to take advantage of the array gain opportunity by accounting for the information that the bases possess regarding the user's angular spread.. The channel model used is derived from electromagnetic propagation principles and serves to furnish a unified setup for observing how the BF and MIMO options emerge and under which conditions an advantage accrues to one or the other. We stay within the (4,4) MIMO realm in our investigations and assume that, although the base transmitter knows the channel statistics, it does not know the channel instantiation.

After developing the propagation model, we review some of the basic scaling and decorrelation properties of the random electric field process**.** Any particular vector of the array element weights is termed an apodization (APOD). Preferred APODs are eigenvectors of the spatial field and their determination is an important concern Associated with each APOD is the corresponding eigenvalue which represents the inherent strength of such a natural field mode. The field correlation integral operator spectrum is its eigenvalue set which we will order putting the strongest first. BF is a special case of MIMO which arises when the APOD spectrum exhibits such a predominant first eigenstate to the extent that other eigenstates are marginalized to the point of being comparatively useless for inclusion as other transmission modes. On-the-other-hand, the significantly advantageous driving of a plurality of eigenstates amounts to MIMO. We will explain asymptotic effects that are evidenced in our numerical study of these spectra, as these characteristics underlie our subsequent BF-MIMO comparisons.

We will also describe the process of *enveloping* a family of channel capacity probability distribution functions, (PDFs). The enveloping amounts to an optimization of APODs toward the goal of capacity maximization at specified outage, or, equivalently, outage minimization at a specified capacity. For simplicity, we first discuss enveloping for MISO. Then we explore enveloping in the context of (4,4) MIMO, while being watchful for where in parameter space MIMO offers a worthwhile advantage over (1,4) BF.

We will investigate elemental antenna array elements with a separation greater than or equal to a 1/2-wavelength spacing to avoid the complication of antenna coupling. The 1/2-wavelength spacing enables substantial phase coherence in transmitted signals. Furthermore, an antenna count of 4, as for 4 transmit antennas in (4,4) MIMO, could, eg, draw on 18 or 36 elemental omni antennas if such were needed to constitute the four high gain transmit antennas along an aperture 4 or 8 decorrelation lengths long respectively. Underlying elemental antennas at $\lambda/2$ spacing spanning a correlated field along an aperture segment are part of the reason for large MIMO and BF capacity gains. Such a plurality of elemental antennas can be seen as heroic. However, arrays of elementary antennae is a *notional design*, likely to be improved through judicious use of directional elements: more advanced antenna design is expected to reduce the number of actively controlled antenna inputs as discussed at the end of this paper.

We will exhibit performance curves for a single MIMO link that show the preference for BF at lower angular spreads, shorter aperture lengths and lower carrier frequencies. On-the-other-hand we will see a preference for MIMO at larger angular spreads, longer aperture lengths and larger carrier frequencies. The preference for MIMO will also depend whether one puts significant value on, say, a 20% enhancement of capacity; if so, then using MIMO can be preferable as we shall see.

The capacities uncovered here may seem large. Indeed, we will we find capacities in bits/symbol of (4,4) MIMO and (1,4) BF to be 5.7 and 3.8 respectively (as compared to 2.6 and 1.5 in previous studies involving widely separated omni antennas at both transmit and receive sites). These results are for a 1.68m aperture length which is modest for MIMO. At twice this aperture length, MIMO (4,4) and BF (1,4) capacities improve to 7.3 and 3.9 respectively.

### B. Relationship to References

References [1-6] relate to the development of the propagation model.

In BF the top eigenmode gets all the power, but for MIMO, one must be careful in powering the transmit modes There is considerable literature exemplified by [7]-[23] investigating the powering of multiple transmit antenna links for MIMO in circumstances where the transmit antennae are, in some way, spatially correlated. Many deal with the ergodic channel case with a capacity maximization objective, but some, like [8,11,12,13,14,18] also deal with the quasistatic channel where capacity outage is of interest. Some deal with both.

Here we are concerned with the quasistatic channel for when limiting to a specified outage probability is important. We will mostly follow pioneering work of Moustakas, Simon and Sengupta [11] and [12-13] by the first two of those authors which includes, (along with work on more advanced methods) their experience with a Monte Carlo approach in addressing BF versus MIMO issues. Our numerical explorations of MIMO focusing on 10% outage will draw on Monte Carlo evaluation for the required capacity PDF enveloping. Reference [12] gives the closed form solution for the MISO case. In reviewing PDF enveloping here we uncover a small outage asymptotic feature for the MISO case. We found reference [24]'s advice for unbiased polytope sampling useful for Monte Carlo sampling of the candidate power vectors. Chalk's penetrating, thorough analysis [25] of the important integral equation that is at the center of our work is another key reference. Chalk's very different communication problem to find the best selection from a certain class of RC pulses for optimal data transmission led him to the same exponential integral kernel that will be the focus of our attention. This integral equation has come to be called the Lalesco-Picard (L-P) equation which traces back over a century to Trajan Lalescos's papers [26-27]: (Emile Picard of Picard iteration fame was his thesis advisor.)

To our knowledge, our investigation of BF (1,4) as compared to MIMO (4,4) is more fundamental and complete here than what has been reported in the literature. We start from propagation fundamentals, under the assumption that the base transmit array has only statistical information regarding the random electrical field along a linear transmit array aperture. We are led to analyze fundamental APOD spectra. The L-P operator's spectra are the APOD spectra that, over parameter space, exhibit various asymptotic tendencies that we will point out. The onset of a MIMO preference stems from an asymptotic effect, at the high end of the APOD eigenvalues to cluster together. Then, driving the top electric field eigenstates we compute 10% outage capacity performance curves for the BF and MIMO techniques. Capacity is expressed as a function of degrees of angular spread, as well as receiver SNR and transmit array aperture length, measured in field decorrelations.

When reference to very early MIMO (4,4) comparison results are made we draw on references [28] and [29].

### II. CORRELATION STRUCTURE OF SIGNALS

In typical cellular scenarios, communications occur from base stations, placed above clutter, to mobiles, usually immersed in clutter. The resulting path loss and correlation was analyzed in [1] for a mobile placed on a line perpendicular to the linear array of antennas ("broadside" direction). Assuming the mobile was in a street orthogonal to the broadside direction, both path loss and correlation at the base were derived and found to be consistent with widely used models (e.g. COST-231-Hata and power angular spectra reported in [2] and [3]). Here we extend that work by generalizing it to mobiles placed off broadside and without the restrictive assumption of a street always running perpendicular to the line connecting mobile and base. Considering propagation from mobile to base, the emitted signal is taken to undergo diffuse scattering [1] in the vicinity of the mobile, generating a field in the region above the mobile, modeled in [1] as spatially white. This field is "visible" from the base, with a signal received at the base that may be evaluated via Huygen's principle. In [1] the signal correlation at the base as a function of horizontal base antenna separation $\mathbf{r}_d$ was found to be related to the two dimensional integral over the power intensity distribution in the horizontal clutter-top region above the mobile, with integration variable $\mathbf{r}'$:

$$\langle v(\mathbf{r})v^*(\mathbf{r}+\mathbf{r}_d)\rangle = |G_z(R_0)|^2 \frac{\lambda^2 \pi}{k^2} e^{ik\mathbf{r}_d \cdot \hat{\mathbf{r}}_0} \int_{-\infty}^{\infty} d\mathbf{r}' I_m(\mathbf{r}') e^{-ik\frac{\mathbf{r}'_\perp \cdot \mathbf{r}_{d\perp}}{R_0}} \quad (1)$$

The above formula involves wavelength, $\lambda$, vertical derivative of the Green's function $G_z$ (defined below), wavenumber $k = 2\pi/\lambda$, horizontal range $R_0$, clutter (building) height $z_c$ and mobile height $z_m$. The factor $e^{ik\mathbf{r}_d \cdot \hat{\mathbf{r}}_0}$ in (1) represents the expected phase difference between two base antennas separated by $\mathbf{r}_d$, while the unit vector along the direct ray from base to mobile is $\hat{\mathbf{r}}_0$. This factor, arising when expanding phase variation across the base array may be interpreted as a "beam-steering" vector. The subscript $\perp$ denotes components perpendicular to $\hat{\mathbf{r}}_0$. The intensity $I_m(\mathbf{r}')$ in (1) is defined here as a circularly symmetric distribution dependent on transmit power $P_t$:

$$I_m(\mathbf{r}_m) = \frac{P_t}{4\pi R_1^2} = \frac{P_t}{4\pi\left[(z_c - z_0)^2 + r'^2\right]}, \ r' \in [0,A] \quad (2)$$

This may be visualized as an open "piazza" of radius $A$, with intensity set to zero outside this circular region.

For flat terrain, the vertical derivative of the Green's function depends on base height $z$ relative to clutter height as [1]:

$$|G_z(R_0)|^2 \approx \frac{z^2}{\lambda^2 R_0^4}. \quad (3)$$

Thus

$$\langle v(\mathbf{r})v^*(\mathbf{r}+\mathbf{r}_d)\rangle = \frac{z^2}{R_0^4}\frac{\pi P_t}{2k^2}e^{ik\mathbf{r}_d\cdot\hat{\mathbf{r}}_0}\int_0^A I_m(r')J_0\left(k\frac{r'r_d}{R_0}\right)r'dr'. \quad (4)$$

Received signal power $P_r = \langle |v(\mathbf{r})|^2\rangle$ may be evaluated by setting $r_d=0$:

$$P_r = \langle |v(\mathbf{r})|^2\rangle = \frac{z^2}{R_0^4}\frac{\pi P_t}{4k^2}\ln\left(1+\frac{A^2}{(z_c-z_0)^2}\right) \quad (5)$$

while (4) may be evaluated asymptotically as $A \to \infty$

$$\langle v(\mathbf{r})v^*(\mathbf{r}+\mathbf{r}_d)\rangle \xrightarrow[A\to\infty]{} P_r e^{ik\mathbf{r}_d\cdot\hat{\mathbf{r}}_0}K_0\left(kr_d(z_c-z_0)/R_0\right) \quad (6)$$
$$\xrightarrow[A\to\infty]{} P_r e^{ik\mathbf{r}_d\cdot\hat{\mathbf{r}}_0}e^{-kr_d(z_c-z_0)/R_0}$$

in agreement with [3]. Numerical evaluations of (4) are compared to the asymptotic expression (6) and the results for the mobile in the street [1] in Figure 1. At zero antenna separation, correlation is the same as path gain, $P_r/P_t$, and may be observed to be about $2\times10^{-14}$, or 137 dB for 2 GHz and 1 km range used to generate Figure 1.

Asymptotic approximation (6), using the modified Bessel function $K_0$ clearly fails at 0 separation, as this function approaches infinity at 0, but this is not much of a concern as (5) applies there. Oscillations observed in the numerical evaluation (4) are due to "ringing" that arises in the abrupt termination of the integral, as implied by the simple piazza model. Overall, the correlation and path loss appear to be only weakly dependent on the shape and size of the scattering region around the mobile, whether it is a circular piazza, as here, or a street as in [1]. We therefore conclude that an exponential correlation model, consistent with (6), (24) in [1] as well as deduced from measurements in [3] is an accurate representation, with a simple phase offset factor $e^{ik\mathbf{r}_d\cdot\hat{\mathbf{r}}_0}$ accounting for off-broadside mobiles.

### III. THE MISO CASE

We first consider the special case of MISO, where M base transmit antennas radiate a signal $s(t)$ meant for reception by a single antenna (mobile) receiver in the presence of noise $n$. The transmitter has the ability to apply complex weights $a_n$. The received signal is

$$r(t) = s(t)\sum_{m=1}^M a_m h_m + n(t). \quad (7)$$

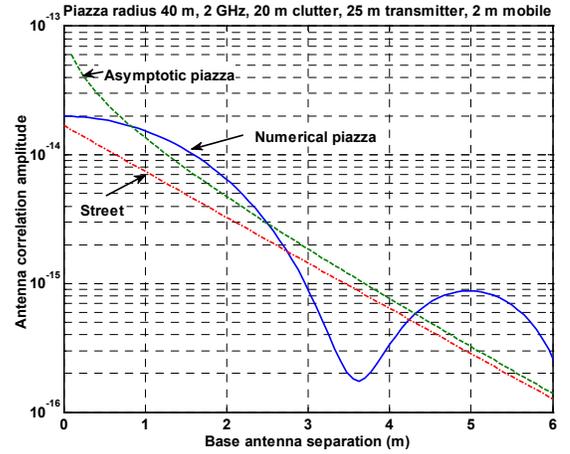

**Figure 1**. Amplitude of channel response correlation coefficient as a function of base antenna correlation, at 1 km range. Values at 0 separation represent path loss (~137 dB at 1 km).

For a given realization of channel coefficients $h_m$, the resulting communication is that of an AWGN channel whose capacity is maximized by maximizing the SNR. We are interested in the case where only the statistics of $h_m$ are known at the transmitter. Within the array, the channel coefficient distribution is taken as complex Gaussian with zero mean and some (average-path loss determined) variance. The complex Gaussian distribution is usually employed to represent the effect of multiple scattered arrivals. The coefficients are correlated with the correlation described in the previous section. It is instructive to also write a continuous aperture version of equation (7):

$$r(t) = s(t)\int_{-A/2}^{A/2} dy\, a(y)h(y) + n(t) \quad (8)$$

The complex weight distribution or "APOD" $a(y)$ is to be optimized. The channel response $h(y)$ is the response occurring in the absence of antennas. It should be noted that considering the received signal as a continuous linear weighting (8) of the spatial channel response is an approximation valid only in the context of antennas that are spaced closer than the coherence scale of the field and further than distances at which inter-antenna electromagnetic coupling becomes prominent. An important example of this is the case of antennas separated by $\lambda/2$ and placed above clutter, where narrow angular spread (several degrees) implies a coherence scale an order of magnitude larger than $\lambda/2$. Placing independently active antennas closer than $\lambda/2$ would necessitate accounting for inter-antenna coupling effects that become important at closer spacings [5]. Keeping in mind such limitations, continuous aperture modeling as in (8) is found to be a useful approximation for insight into the performance of antenna arrays. For single stream communication in the presence of AWGN of variance $\sigma^2$, the capacity

$$C = \log_2\left(1 + \frac{P_r}{\sigma^2}\right) \quad (9)$$

is a monotonic function of the (desired) received signal power:

$$P_r = P_t \int_{-A/2}^{A/2} dy\, a(y)h(y) \int_{-A/2}^{A/2} dy'\, a^*(y')h^*(y') \quad (10)$$

where the transmitted power $P_t = \langle |s(t)|^2 \rangle$. Taking the view that temporal signal variations in a typical wireless channel are slow relative to coding block duration, capacity (9) may be viewed as a random variable. As the dependence of capacity on received power is monotonic, capacity is maximized by maximizing received power (10), subject to a constraint on total radiated power $P_t$, leading to the normalization

$$\int_{-A/2}^{A/2} dy\, |a(y)|^2 = 1. \quad (11)$$

As the channel coefficients $h$ are Gaussian-distributed, the received signal (8) is also. The power of such a signal is maximized at all outages by maximizing its average power:

$$P_r = P_t \int_{-A/2}^{A/2} dy'\, a(y') \int_{-A/2}^{A/2} dy''\, a^*(y'')R(y'-y'') \quad (12)$$

where the channel correlation is

$$\langle h(y)h^*(y+y_d) \rangle = R(y_d) \sim e^{-\alpha|y_d|}. \quad (13)$$

The exponential form of the correlation was justified in the previous section. By the Cauchy-Schwartz inequality, average received power (12) is optimized when the APOD $a(y)$ is an eigenstate of the channel correlation function, with a corresponding eigenvalue $\nu$

$$\nu a^*(y) = \int_{-A/2}^{A/2} dy'\, a^*(y')R(y-y'). \quad (14)$$

For discrete antennas, the equivalent statement is

$$\mathbf{R}\mathbf{a}^* = \nu \mathbf{a}^*. \quad (15)$$

From (12) it follows that if the $k^{th}$ eigenstate is excited, the received power is

$$P_r = \nu_k P_t. \quad (16)$$

The correlation function $R$ viewed as an operator has a discrete spectrum: Average received power, and thus capacity, are maximized by choosing as APOD the eigenstate $a(y)$ possessing the maximum eigenvalue $\nu_{max}$. This result is valid only if the transmit array is allowed to radiate a single signal stream. Later it will be shown that outage capacities may be improved through radiating different signals on different eigenstates.

### IV. SPECIAL APODIZATIONS AS EIGENSTATES

The random electric field process, $G(s)$, along a linear finite aperture parameterized by s, is given by the following expression, which is its, so called, Karhunen-Loeve (K-L) expansion, (see for example [30]):

$$G(s) = \sum_{1}^{K} \nu_k^{1/2} z_k \phi_k(s). \quad (17)$$

We will review properties of this well known type of expansion in our context.

The $z_k$ are complex iid zero mean Gaussian random variables of variance one. In our application the $\phi_k(s)$ are called APOD functions. They are the eigenfunctions, also termed eigenstates, of the integral operator (14) which has corresponding eigenvalues given by the $\nu_k$ which are taken to be listed in decreasing order. The sequence of $\nu_k$ are called the spectral values: they serve as variances in that the $(\nu_k)^{1/2} z_k$ products are statistically independent Gaussian random variables that have variances $\nu_k$.

The first term in the K-L summation (17) is the best average mean square finite energy approximation to $G(s)$ over the length of the aperture. The average mean square error uses a product space integral. This integration is expressed as $E\int$ amounting to taking the integral along the finite aperture as well as the underlying probability space, where $E$ stands for expected value. The first two terms amount to the minimum average mean square error for a two term expansion, and so on for three terms, etc.

We will be interested in the discrete aperture finite $K$ case and the continuous case of infinite $K$. In the discrete case the integral is just a sum and the integral operator is the $K \times K$ matrix formed by an equispaced sampling of the kernel function. In this case, we require the resolution of equispaced samples to be no finer that a half-wavelength: going below that would soon involve the onset of antenna coupling which brings in electromagnetic theory that is beyond the scope of this paper. In the continuous case, the integral is an ordinary integral along the aperture. The infinite aperture continuous case limit is also of interest.

The trace of the autocorrelation operator, whether discrete or continuous, is equal to the sum of its eigenvalues. This will be evidenced in the figures in the next section.

For apertures that are intervals, the $\phi_k$ comprise a Fourier system, ie sines and cosines. This can be seen by differentiating the integral equation to obtain a differential equation that is satisfied by cosines (and sines) of a certain transcendental equation (or its nonidentical twin equation for sines). The transcendental pair gives the natural frequencies of the system. Reference [25], for which the same integral equation kernel arose in a data pulse optimization application, is an excellent reference for development of this specific Fourier eigenstructure. In the K-L expansion the $z_k \phi_k$ are biorthogonal, ie.

$$\int_{-A/2}^{A/2} dx \phi_i(x) \phi_j^*(x) = \delta_{ij}, \quad \langle z_i z_j^* \rangle = \delta_{ij}. \quad (18)$$

where the Kroneker delta appears on both right hand sides. In our application, when the user azimuth angle is zero the $\phi_k(x)$ are real and * which denotes complex conjugate is not needed as it is when the user is located off angle. The K-L expansions here are "complete" meaning that average mean squared error of the full expansion is zero in the finite case as it is in the limit of mean square error in the infinite case. The eigenvalues, $v_k$, are distinct.

## V. SCALINGS FROM THE DERIVED ANGULAR SPREAD MODEL

Empirical measurements of the statistical behavior of the radio fields are usually reported as power angular spectra (PAS), which is a Fourier transform of the spatial correlation (Wiener-Khinchin). Taking the field correlation at the base to be exponential as in (6) and (13), the resulting angular power spectrum is given by

$$P(k_y) \approx P_r \frac{2\alpha}{\alpha^2 + (k_y - k_{y0})^2} \quad (19)$$

where the spatial frequency along the aperture $k_y$ is related to the azimuth angle $\phi$, measured from the normal to the base array, by $k_y = k \sin \phi$, the offset frequency $k_{y0} = k \sin \phi_0 = k \mathbf{r_d} \cdot \hat{\mathbf{r}}_0$ is the phase ramp across the linear base array aperture arising when the mobile is off-broadside. The correlation is parameterized (6) by $\alpha = k(z_c - z_0)/R_0$. It is in turn related to the 3 dB full-beamwidth $\phi_{3dB}$ by

$$\alpha = (2\pi f/c)\phi_{3dB}/2 = (\pi/\lambda)\phi_{3dB}$$
$$= 1/(Decorrelation\ distance) \quad (20)$$

We see that the decorrelation length varies:
- inversely with carrier frequency, (directly with λ)
- inversely with angular spread.

As is typical, we take the onset of antenna array element coupling in heavy scattering to be $\lambda/2$. So the number of half $\lambda$s that are in one aperture decorrelation distance is

$$(decorrelation\ distance)/(\lambda/2) = (1/\alpha)/(\lambda/2)$$
$$= (2/(2\pi\phi_{3dB}/\lambda))/(\lambda/2) = 2/(\pi\phi_{3dB}). \quad (21)$$

The *rms* angular spread $\sigma$ is reported by Pedersen, et al, [4] as ranging from $2^o$ to $10^o$ in the probability range [0.2,0.8] for a tall (32 m) tower in Aarhus, Denmark. Measured power angular spectra are found to be reasonably well represented by the exponential form $P(\phi) = e^{-|\phi|\sqrt{2}/\sigma}$. Later, Andersen and Pedersen [3] report that the same data is even better represented by the Cauchy-Lorenz distribution. $P(\phi) = 2a/(a^2 + k^2 \sin^2 \phi)$. with $a$ interpreted as the 3 dB full-width measure of angle spread. This is related to previously reported measure as $a = k \sin \phi_{3dB}/2$, with $\phi_{3dB} = \sqrt{2}\sigma ln\ 2$. It may be observed in Figure 2 that the two functional forms agree closely up to -15 dB. In our work we represent the power angular spectra also as Cauchy–Lorenz, in line with both the latest empirical reporting as well as theoretical modeling.

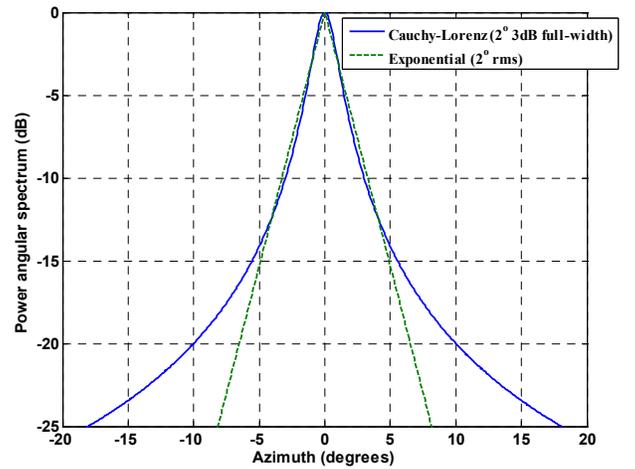

**Figure 2.** Power angular spectra

The range of 3 dB full-widths corresponding to the reported range of $\sigma$ is $\phi_{3dB} \in [2^o, 10^o]$. Theoretical considerations suggest $\phi_{3dB} = (z_c - z_0)/x$, where $x$ is the range, $z_c$ clutter (building) height and $z_0$ mobile height. The angle spread is thus predicted to decrease with range.

We choose the value of $\phi_{3dB} = 2°$ as representative of points further away from the base. From a MIMO point of view this is conservative as it allows only modest spatial independence over modest apertures. Mean values of full width $\phi_{3dB}$ suggested by the measurements [4] are 8°. Furthermore, measurements [4] collected at other urban areas and with lower BS heights (still above rooftop), report a median $\phi_{3dB}$ of 10°. It may be said, therefore, that the onset of better MIMO performance would occur for most cases in urban areas for much smaller apertures.

The first, second and 10th eigenmodes and corresponding power angular spectra are plotted in Figure 3. The power spectra are scaled by the corresponding eigenvalue so as to give an indication of the relative power of the mode. The modes are seen to be cosines/sines whose spatial frequency increases with the mode number, while the eigenvalue strength decreases monotonically. It may be observed that the first "fundamental" mode is a portion of the single period of a cosine. It is seen that the corresponding power angular spectrum is a narrow beam aimed at the center of the scattering region around the mobile (0° in azimuth for this broadside case). Higher order modes have beams consisting of two lobes, aimed at either side of the scattering region center. As the mode number increases, the lobes are shifted further away from center. The beam strength correspondingly decreases as the lobes are aimed at weaker parts of the scattering region around the mobile. With increasing angle spread, the size of the scattering region increases, making higher order modes more significant despite being aimed off-center.

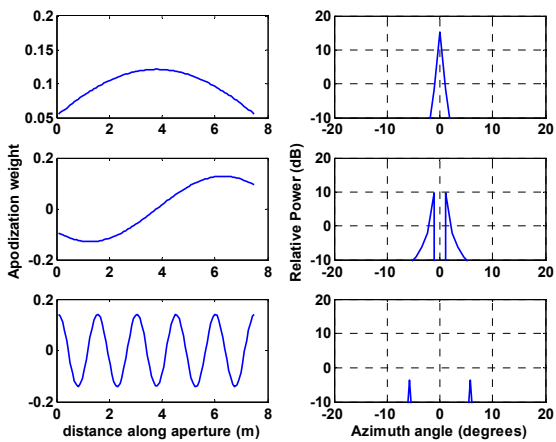

**Figure 3.** APOD eigenstates (left) and corresponding power angular spectra (right). Top to bottom are modes 1, 2 and 10.

## VI. APODIZATION SPECTRA AND ASYMPTOTES

In this section, we look at APOD spectra and some of their asymptotic trends. The abscissas in the four figures that follow feature the number of antennae as well as the number of antennae per decorrelation length in a linear aperture. The ordinates present a vertical plot of the eigenvalues at the abscissa values. In the figures we will see many cases where it is apparent that the operator trace is the eigenvalue sum as it must be.

In Figure 4 the abscissa is the number of equispaced antennae along a one decorrelation length aperture. We witness the topmost eigenvalue growing essentially linearly as more and more antennas are placed within the fixed one decorrelation length aperture up to a maximum of 18 antennas. Beamforming is associated with this principal eigenvalue which is seen to depart from the other eigenvalues by an increasing amount as the number of antennas is increased over the length of one decorrelation.

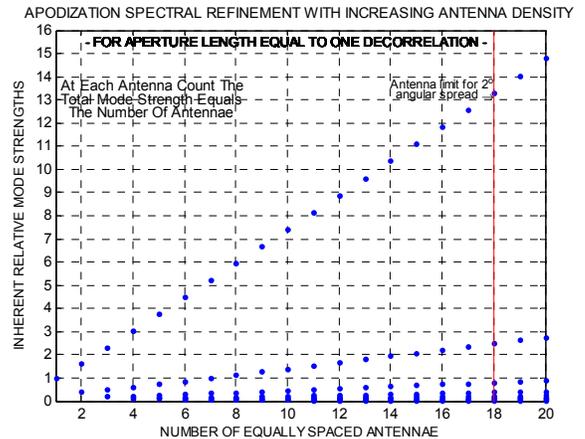

**Figure 4**. The growing departure of the BF eigenvalue from all the lesser eigenvalues due to progressively tighter antenna packings along a linear aperture one decorrelation length long. Tighter antenna packings within a coherence length enables very significant "antenna gain" for BF.

A maximum antenna count of 18 indicated by the red line, would occur, for example, for a 2 GHz carrier frequency and 2° angle spread, where the number of decorrelations within the aperture is 18. In such a case one must stop at 18 antennas to adhere to the proscription of neighboring antennas being no closer than a half-wavelength.

$J_K$, the K by K matrix with entries of all ones has rank one with the nonzero eigenvalue equal to K. The linear growth of its top eigenvalue with K is akin to the standout performance exhibited in Figure 4 of the top eigenvalue as compared to all the weaker ones. With the equispaced antenna spacing along the aperture, the $K \times K$ autocorrelation matrix of sampled $\exp(-\alpha|d|)$, does not quite lead to the constant entry $J_K$ matrix, for an angle spread of 2°. However, the 2° autocorrelation decays so slowly as to give a $K \times K$ matrix with somewhat similar linear eigenvalue behavior. The nonzero geometric decay in Figure 4 reaches an eigenvalue of 13.2 for 18 antennas falling short of the eigenvalue of 18 for $J_{18}$. Note that the secondary, tertiary eigenvalues, etc, also show evidence of linear growth with the number of antennas.

In Figure 5 we see the principal (top) eigenvalue saturating near 2.1. Here there is only one antenna per decorrelation length and the abscissa goes out to 30 decorrelations. The decorrelation due to widely spaced antennas underlies the great reduction of array gain compared to the previous case Essentially linear antenna gain of the top APODs was evidenced in Figure 4, instead, here in Figure 5, we see the top APODs transitioning to express a clear onset of saturation of antenna gain. Next, we will observe another example of saturation but at a significantly higher level.

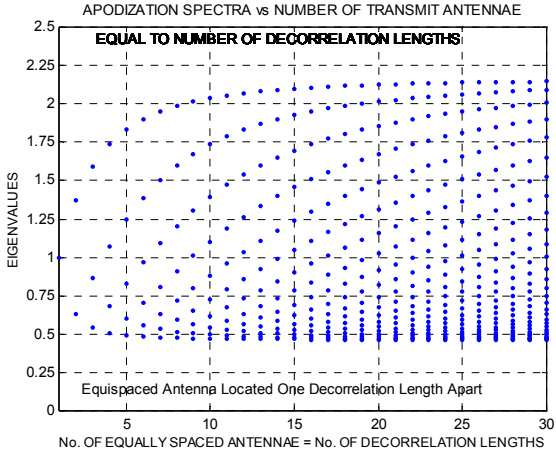

**Figure** 5 Illustration of greater clustering together of the larger eigenvalues due to a progressive increase of the number of antennae and a proportionate increase in aperture length so that there is one antenna per decorrelation length.

Fig. 6, like Fig. 5 exhibits top eigenvalue saturation because of the large number of decorrelations along the aperture compared to Fig. 4 which just assumed one decorrelation. Unlike Fig. 5, Fig. 6 has the maximum antenna gain permitted since it assumes the closest antenna spacing allowed per decorrelation length. This close antenna spacing gives very substantial antenna gain, so we the saturation is nearing 36 not 2.2 as in Fig. 5. This is due to combining the key features of environment for Fig. 4, namely, a large number of closely spaced antennas within a decorrelation with a key feature of the environment for Fig. 5, namely, a large number of decorrelations. So we observe both large eigenvalues and their clustering.

**Theorem 1.**
*In the limit of infinite aperture, apodization spectra become sinusoidal for an arbitrary (symmetric) power angular spectrum*
*Proof:*
In the limit of increasing array aperture $A$, (14) becomes

$$va(y) = \int_{-\infty}^{\infty} dy' a(y') R(y - y') \quad (22)$$

the right side of which is an ordinary convolution integral with a space-invariant kernel $R$. This eigenvalue equation may be solved by taking the Fourier transforms of both sides:

$$v\mathrm{A}(k_y) = \mathrm{A}(k_y) S(k_y), \text{ where}$$
$$\mathrm{A}(k_y) = F\{a\}, \quad S(k_y) = F\{R\}. \quad (23)$$

Solutions of (23) are thus

$$v_n = S(k_y),$$
$$\mathrm{A}(k_y) = \delta(k_y - k_{yn}) \Rightarrow a(y) = e^{ik_{yn}y} \quad (24)$$

For symmetric angular power spectra, such as (19) (at broadside, where $k_{y0}=0$), the solutions are degenerate and occur in symmetric pairs of exponentials, corresponding to sines and cosines. ∎

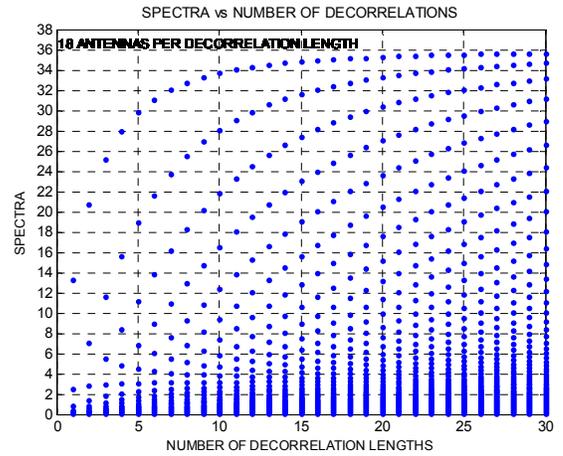

**Figure 6**. As the number of decorrelations increase, enormous eigenvalues for the top APODs occur and these larger eigenvalues are seen clustering together. We also see the onset of spectral transition from an obviously predominant BF eigenvalue at a correlation length of unity to the clustering of a plurality of the larger eigenvalues favoring MIMO at the large correlation lengths (abscissa values).

**Theorem 2.** Given power angular spectrum (19), in the limit of infinite aperture and with elementary antennas separated by λ/2, the maximum eigenvalue of (22), with corresponding to the beamforming gain saturates at 2/α. Furthermore, the number of secondary modes with eigenvalues in the arbitrarily small neighborhood of the 2/α increases linearly with aperture length.

*Proof*: The eigenvalues of (22) satisfy a transcendental equation [25], and form a discrete set of values corresponding to spatial frequencies $k_y = k_n$, with $k_n$ scaling with aperture length $A$ as $1/A$. As $A$ increases, any particular $k_n$ approaches 0 as $1/A$.

To obtain the limiting beamforming gain (24) for the fundamantal mode in the macrocellular case,

$$v = S(k_y) = \frac{2\alpha}{\alpha^2 + k_y^2}, v \xrightarrow[k_y \to 0]{} v_{\max} = \frac{2}{\alpha} = \frac{2R_0}{k(z_c - z_o)} \quad (25)$$
∎

For discrete arrays, with $\Delta y = \lambda/2$ spacing, the corresponding gain limit is $v_{max}/\Delta y \approx 36$ (15 dB), corresponding favorably to the top asymptote in Figure 6.

In Figure 7 we look again at the spectra of eigenvalues, but for a case where there are only 18 equispaced antennas over the entire decorrelation length. Then there is a considerable degree of decorrelation among the 18 transmit antennas for the longer apertures. We see clear evidence of the approach of all the eigenvalues converging to 1, as would be the case for the asymptote of complete decorrelation between each distinct antennas pair. This is because, with such increasing decorrelation, the sampled decorrelation matrices are getting closer to the identity matrix all of whose eigenvalues are 1.

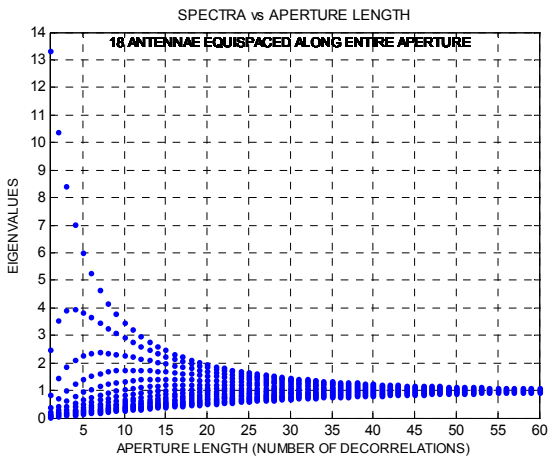

**Figure 7** Evidence, that for apertures growing progressively larger in decorrelation lengths, with a proportionate lessening density of antennae within each decorrelation length, the entire spectrum collapses to unity.

### VII. LOWER ENVELOPING IN A MISO CASE

Here we illustrate the simple numerical determination of the best way to distribute power between two APODs in a multiple input single output (MISO) case. This is to lead up to our main interest which will be in much more complex higher dimensional situations. Here we address address a hypothetical case where, from the K-L expansion we have that the strongest two APODs among the candidates has eigenvalues of 5.24 and 3.63, like we see in Figure 8, when the number of decorrelations is 6. Assume these top two APODs and a single receive antenna for the (2,1) MISO system.

We want to maximize the argument of a capacity function that has the form

$$Log_2(1 + \lambda_1 5.24\, E_1 + \lambda_2 3.63 E_2) \qquad (26)$$

where $\lambda_1, \lambda_2 \geq 0$, $\lambda_1 + \lambda_2 = 1$. $E_1$ and $E_2$ are independently distributed exponential random variables of mean 1 corresponding to the modulus squared of zero mean, unit variance complex Gaussian variates. All the transmitter knows are the APOD eigenvalues and their probability distributions, but not their outcomes.

The reason that it is advisable to look beyond just using the strongest APOD is that the second APOD offers a statistically independent choice. Involving the second APOD offers a hedge in cases when the strongest variance APOD has a weak outcome. We will look into adequately resolved candidate splits of available power and choose the best split for maximizing the capacity at minimum outage. We consider all power splits with a 1% gradation in split values, always selecting that split which gives minimum outage. This amounts to looking at a candidate family of 101 capacity PDFs and selecting the best PDF at each outage. So we pick the lower envelope over the candidate PDF family, which is itself a PDF, namely that special PDF having the feature of providing the lowest outage.

Figure 8 exhibits the effect of the power splits using, the top two APOD strengths as for a decorrelation length of 6 in Figure 7. Starting from a point on the weaker magenta PDF, as one moves vertically down to the next lower blue dot, another 1% of the stronger red PDF is mixed at the expense of a 1% reduction of the weaker PDF. Moving further and further down, one passes the red PDF and gets to the mix yielding the lowest PDF shown here in black. After the black PDF is reached, further mixing to add more and more of the stronger PDF ultimately gives the stronger red PDF which is inferior outagewise to the lower envelope black PDF.

We focus here on the 10%-tile. Consider an imaginary vertical line erected at an abscissa value which pierces the lower envelope at 0.1 (ie 10%) outage. We see that the optimal mixing of the two PDFs was worthwhile, as we get 1.12 bits per cycle. This is contrasted with getting 0.62 bits/cycle for using just the stronger channel. (The weaker (magenta) channel choice gives only about 0.52 bits/cycle). From the numerical values it is seen that forcing a 50/50 choice is not a bad choice for this case. It gives 1.11 bits/cycle in contrast to 1.12 for the optimum 48/52 split. In a context of this specific example we just got a hint of a general MISO asymptotic result, expressed in the following theorem.

**Theorem 3. Equal transmit powers amongst modes maximizes MISO capacity in the asymptotic limit of low outage.**

We elaborate this theorem in more detail: Given a set of *independent* exponentially distributed natural mode strengths, such as arising in the context of a MISO AWGN channel with independent zero-mean complex Gaussian channel coefficients, with transmitter only knowing the average power of each, the MISO capacity is optimized in the limit of extremely low outage, through equal distribution of total available transmit power amongst all modes.

*Proof.* See the Appendix.

Related results, which rely on stereotypical small outage properties, can also be derived for the MIMO case with an arbitrary finite number of inputs and outputs, but we will not do so in this paper.

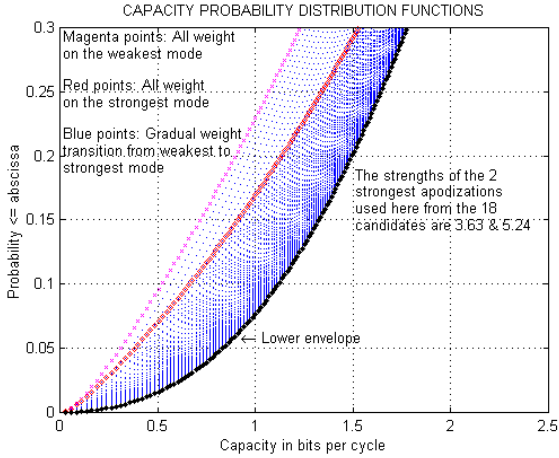

**Figure 8**. Lower Enveloping of 101 candidate capacity PDFs formed from splitting the available power between the two strongest of 18 candidate APODs. Optimum mixing of the strongest (red) and next strongest (magenta) PDFs yields the lower envelope black PDF, offering significant improvement over just using the strongest PDF.

## VIII. DISTRIBUTING POWER OVER MODES TOWARDS MAXIMIZING MIMO OUTAGE RATE

References [28] and [29] dealt with MIMO (m,m) systems where all m inputs had equal variances, and so, for low outage (say, eg, 10%), equal power drivers were tacitly assumed. Here we have a more general situation in that the variances are typically not equal. We will again proceed under the assumption of driving the top APODs with statistically independent drivers, but the powers of the drivers must be estimated with the goal of curtailing outage. The independence assumption disallows any correlation among the drivers. Indeed, we are interested here in only low outage applications, typically (4,4) systems with a 10% outage limitation, for which we have no motivation in including the complication of the option of correlated random driving processes.[*]

So how do we estimate the optimal powers of the statistically independent drivers? Some options are: a) Do a random search or b) Transform to the equal variance case and use equal power at the expense of getting a bound. (Later we mention how that can be done.) or c) Solve with optimization theory.

---

[*] The high outage realm is different, because there it can be useful to "bet" on the rare event that significant spatial coherence among random channel components occurs with correlated drivers actually improving reception.

We will take the random search approach. How does one avoid biased searching? How can we be cautious not to miss the "needle in a haystack"? We address this next.

### A. Unbiased Sampling of the k Powers Summing to P

To start, the power P will be normalized to one and we will follow the advice of [24] to get unbiased samples. Namely, take K-1 *iid* samples that are uniform over [0,1]. Consider them placed in [0,1] where $x_1, x_2, \ldots x_{k-1}$ lists the samples largest to smallest. Illustrating the general case with the case K = 4, we get the needed samples from $\lambda_1 = x_1$, $\lambda_2 = x_2 - x_1$, $\lambda_3 = x_3 - x_2$ and $\lambda_4 = 1 - x_4$.

### B. Monte Carlo Learning Algorithm for Distributing P Over Modes to Maximize Rate at a Specified Outage

Here we assume (m,n) MIMO with fixed SNR, allowed outage and total power P for driving all active APODs. We will use **H** to denote the n by m matrix whose entries are 0 mean complex Gaussian random variables of variance 1. The SNR parameter $\rho$ is the expected SNR at a single receive antenna receiving a signal launched from one omni transmit antenna. The variances of the APOD eigenstates are known to the transmitter, but not the random outcomes.

In the canonical basis corresponding to elemental antennas, the channel matrix with separable receiver/transmitter correlations may be written as

$$\mathbf{H} = \mathbf{\Psi}_R^{1/2} \mathbf{H}_{iid} \mathbf{\Psi}_T^{1/2} \qquad (27)$$

where the SVD decomposition of $\mathbf{\Psi}_T$ is $\mathbf{\Psi}_T = \mathbf{\Theta}_T \mathbf{S}^2 \mathbf{\Theta}_T^\dagger$ with unitary $\mathbf{\Theta}_T$ and diagonal $\mathbf{S}$ containing the square roots of the known variances (modal strengths). Thus

$$\mathbf{H} = \mathbf{\Psi}_R^{1/2} \mathbf{H}_{iid} \mathbf{\Theta}_T \mathbf{S} \mathbf{\Theta}_T^\dagger = \mathbf{\Psi}_R^{1/2} \mathbf{H}'_{iid} \mathbf{S} \mathbf{\Theta}_T^\dagger \qquad (28)$$

where $\mathbf{H}'_{iid} = \mathbf{H}_{iid} \mathbf{\Theta}_T$. The received signal is $\mathbf{y} = \mathbf{H}\mathbf{x} + \mathbf{n}$, similar to (7), more conveniently written with excitation $\mathbf{x}' = \mathbf{\Theta}_T^\dagger \mathbf{x}$ in the basis of the field correlation at the transmitter $\mathbf{\Theta}_T$:

$$\mathbf{y} = \mathbf{H}\mathbf{x} + \mathbf{n} = \mathbf{\Psi}_R^{1/2} \mathbf{H}'_{iid} \mathbf{S}\mathbf{x}' + \mathbf{n}. \qquad (29)$$

At a specified outage level, we would ideally like to find the diagonal matrix, $\mathbf{\Lambda}$, whose $\lambda_k$ diagonal entries, representing powers of $\mathbf{x}'$, maximize

$$log_2 \left| \mathbf{I} + \rho \mathbf{H}_{iid} \mathbf{S} \mathbf{\Lambda} \mathbf{S}^\dagger \mathbf{H}_{iid}^\dagger \right|. \qquad (30)$$

with $trace(\Lambda) = P$. In the diagonal matrices $\mathbf{S}$ and $\Lambda$ we arrange both the variances $v_k^{1/2}$ and the $\lambda_k$, respectively, in decreasing order down the diagonals. The matrix $\mathbf{H}_{iid}$ has its entries distributed as zero mean, unit power complex Gaussians, appropriate for representing the channel coefficients between each eigenmode (in the K-L expansion (17) at the base) and each receive antenna at the mobile where rich scattering is assumed, leading to uncorrelated Rayleigh spatial fading. The average path loss effects together with shadow fading are absorbed into the SNR $\rho$ as is the unnormalized value of P and receiver noise level. The normalization stemmed from the absorption of $\rho$ in each diagonal term in the product involving $S$ and its transpose conjugate which straddle $\Lambda$ in the LogDet expression. This accounts for the normalization of powers that allows working instead with the K dimensional vector $\lambda$ whose nonnegative entries $\lambda_1, \lambda_2, ... \lambda_K$ sum to one. Our goal is to find the lower enveloping PDF for maximizing rate at the specified outage.

### C. Monte Carlo Algorithm to Estimate the Maximizing Rate at Specified Outage.

Assume the candidate $\lambda = \lambda_1, \lambda_2, ... \lambda_K$ ordered strongest to weakest are uniformly searched over the polytope

$$1 > \lambda_1 > \lambda_2 > ... > \lambda_K > 0. \quad (31)$$

For each candidate, matrices of m by n i.i.d Gaussian channel characteristics $\mathbf{H}_{iid}$ are independently generated and the outage capacity produced by the candidate estimated. Eg, for each of 1,000 candidate $\lambda$, 10,000 $\mathbf{H}_{iid}$ are generated and the record $\lambda$ for maximizing the $100\% \times (1000/10000) = 10\%$-tile LogDet is kept along with the associated record rate. To improve outage capacity accuracy, the process can be iterated. As with the insertion of magnifying lenses, one can also iterate over smaller and smaller polytopes about a solution that is observed to be forming.

### D. Equalizing Transmitted Powers

We next look into the power penalty levied on an unequal APOD variance system so as to enable it to transmit as if it had equal APOD variances. For this, consider the following pair of systems. Two systems are given the same amount of power $P$ to distribute over their $K$ APODs The *uncommon variance* system has all different APOD variances $v_1, v_2, ...v_K$. The *common (also here called constant or equal) variance* system has all equal APOD variances $u_1 = u_2 ... = u_K$. The two systems share the same variance sum, so that

$$v_1 + v_2 + ... + v_K = u_1 + u_2 ... + u_K \quad (29)$$

The objective is to transmit equal power out of each of the uncommon modes. To accomplish this, the $K$ driving APOD powers must clearly be

$$P_k = (P/v_k)/(\sum_{\kappa=1}^{K} 1/v_\kappa) \quad k = 1, 2, ... K. \quad (30)$$

so the $K$ equal output transmit powers are

$$P_k = P/(\sum_{\kappa=1}^{K} 1/v_\kappa), \; k = 1, 2, ... K, \quad (31).$$

which total to $PK/(\sum_{\kappa=1}^{K} 1/v_\kappa)$.

For the common variance system, the driving powers are all $P/K$ and the equal output powers are all $Pu_k/K$. So the total output power is $P \times Avg\{v_k\}$. Consequently, the ratio of the total common system output power to the total uncommon system output power is

$$P \times Avg\{v_k\} / (PK/\sum_{1}^{K} v_k^{-1}) = Avg\{v_k\} \times Avg\{v_k^{-1}\} > 1. \quad (32)$$

This last inequality is a straightforward exercise. In dB, the penalty for the additional power relative to a constant $v_k$ system is, therefore,

$$10 \times log_{10}[\, Avg\{v_k\} \times Avg\{v_k^{-1}\}\,]. \quad (33)$$

In the MISO example, in VIII, where the eigenvalues are 5.24 and 3.63, the penalty is only 0.146 dB.

Another way the need to equalize arises is when one adheres to the limit on total power, distributing it among the APODs so that they radiates equal power. This lowers the bound on capacity for a given outage, since, while one is adhering to the given power one is constraining the distribution over the driven APODs to force them to radiate equal power. The advantage of settling for a bound is that one does not face the power optimization.

Say that, p is the given power, and $p_k$ the power for the $k^{th}$ driven APOD. The formula for $p_k$ is clearly

$$p_k = v_k^{-1}(p / \sum_{1}^{K} v_k^{-1}). \quad (34)$$

For the two equalization options discussed, one can optimize K rather than fix it to a certain value. So with the $v_k$ listed in decreasing order, one can explore where to truncate the sum to get maximum capacity under specified outage for these equalization procedures.

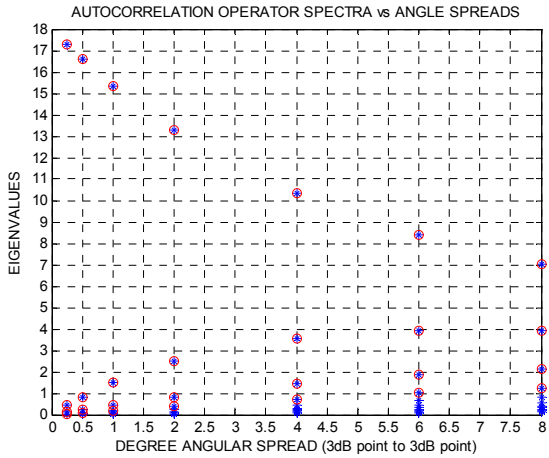

**Figure 9**. Relative APOD strengths as a function of angular spread. At each spread value, most of the 18 eigenvalues plotted are very small, and, so, they visually overlap on the plot: the topmost 4 of the 18 eigenvalues are circled in red. For each angle spread value, the beamforming eigenvalue is the very topmost one.

## IX. COMPUTER EXPERIMENTS EXHIBITING (4,4) MIMO BETTER THAN BF (1,4)

### A. BF (1,4) and MIMO (4,4) Capacities and the Onset of Spectral Collapse Toward Unity

We look at downlink (4,4) MIMO and BF in the form of a (1,4) system and see how they compare. Our investigations will include different values of angular spread. All four receive antennas are assumed to be omni antennas located in clutter. We begin with a look at the top four APODs to see how they fare as a function of angular spread.

Fig. 9 plots, in blue, the 18 eigenvalues as a function of angular spread for $0.25°$, $0.5°$ $1°$, $2°$, $4°$, $6°$ and $8°$. Before we get to $1°$ spread it is hard to observe distinct red points. The overwhelming majority of the 18 eigenvalues are close together from $1°$ out to $8°$. However, at $4°$ to $8°$ the 4 topmost eigenvalues have moved some toward mutual parity due to the increased decorrelation.

As angle spread increases, decreased coherence reduces the strength of the dominant eigenstate (less BF gain) while increasing parity among modes sets in. In the large decorrelation limit, the entire spectrum collapses to unity as is clearly exhibited in the companion Fig. 7 which does not exhibit the sub $2°$ spreads.

In Figure 9, the ordinate value in degrees is equal to two times the aperature length expressed in decorrelation distances. The same is true for the ordinate in Figure 10 which we discuss next as we contrast how MIMO (4,4) and BF (1,4) perform under similar circumstances.

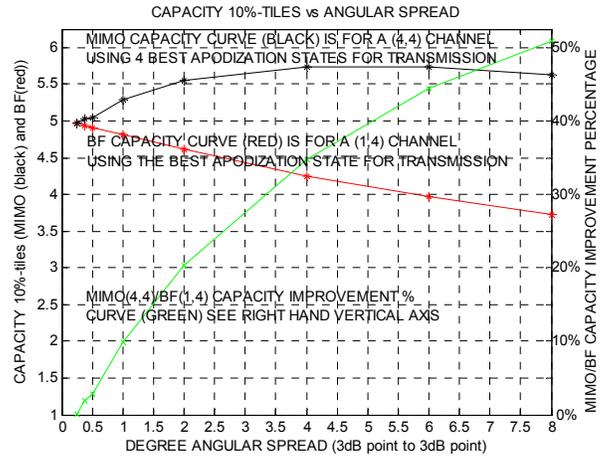

**Figure 10**. Capacity 10%-tile improvement of MIMO over BF for $0.25°$ to $8°$ angular spread. The percent improvement (green), growing to about 51% is quantified by the right hand vertical scale. These curves are for 0 dB SNR.

Toward the top of Figure 10, for 0 dB SNR, we see the 10%-tile allowed capacity outage performance of both (4,4) MIMO (black) and (1,4) BF (red). The latter uses only the top strength APOD. The powers assigned to the top 4 APODs used in MIMO are determined through the numerical Monte Carlo enveloping procedure described above. The two curves do not cross at $0.25°$ since BF is a constrained form of (4,4) where 3 nonprincipal APODs get zero power. The reason that at $0.25°$ the difference between MIMO and BF is hardly discernible is due to the overwhelming strength of the principal eigenstate. As one would anticipate from the spectral plots in Figure 9, the strong emergence of considerably better performing MIMO over BF starts to be exhibited as we move to the right. The relative ratio of the 10%-tile capacities of MIMO over BF is displayed by the green curve which uses the scale at the right. At 0 dB, we see an advantage of MIMO over BF of about 44% at $6°$ angular spread and about 51% at $8°$ spread.

Observe that the (4,4) MIMO curve near the top of the figure rises and then begins decreasing as angular spread increases. This is because, as angular spread increases, antenna gain of each of the four eigenmodes is eroding as those four eigenmodes transition to 4 modes of equal strength as in the (4,4) *iid* matrix channel which does not exhibit an antenna gain. Indeed, the *iid* channel case is devoid of correlation which is the very source of antenna gain. This suggests further improvement with higher order MIMO, say, for a start, (5,5).

### B. Capacity versus Decorrelation for the Tightest Packing of Elemental Antennas

Capacity curves for (4,4) MIMO and BF corresponding to Figure 6 are shown in Figure 11. The gains of MIMO over beamforming range from 21% at the low, single decorrelation end, up to 124% at the very large aperture length of 30 decorellations. At six decorrelations, the gain curve, already well experiencing the onset of saturation gives a factor of 2 gain (ie, 100% improvement).

The (4,4) equal strength lower bound, mentioned earlier in is shown in red and asymptotes, as expected, to merge with the black curve for the longer decorrelations. This lower bound avoids the power optimization step.

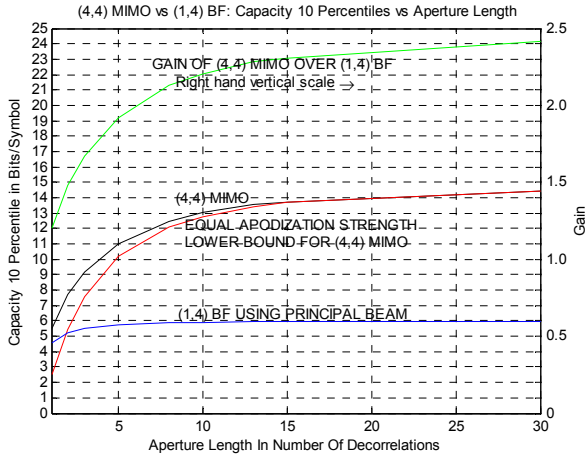

**Figure 11**. Corresponding to the spectral plots in Figure 6, capacities are shown for power optimized (4,4) MIMO along with BF (1,4). The red curve shows the closeness of the (4,4) lower bound at the longer aperture lengths. All three curves are for 0 dB SNR and they assume the tightest $\lambda/2$ antenna spacing all along the aperture.

### C. A Closer Look at the Narrow Angular Spread of 2°

For 2° spread, Fig. 12 depicts the improvements afforded by (4,4) and (2,4) MIMO over (1,4) BF as a function of SNR which ranges from -6dB to +21dB. One decorrelation length and the maximum number of elemental antennas allowed at ½λ spacing within one decorrelation length, namely 18 for a 2 GHz carrier are assumed. The 10%-tile capacity gains of (4,4) and (2,4) MIMO over (1,4) BF are shown using systems that have the power optimized at each SNR for the top 4, 2, and 1 APODs respectively.

The (4,4) equal strength lower bound, that was mentioned earlier in is shown in red and it merges asymptotically with the black curve for the longer decorrelations. This lower bound avoids the power optimization step.

The top four of the 18 eigenvalues are [13.313 2.495 0.822 0.394] and while the base knows these values, it does not know the channel instantiations. Reaffirming what we saw in Figure 12, the gain at 0 dB is 21%. Up to 0 dB, we see that (2,4) MIMO performs almost the same as (4,4) due to the two weakest eigenvalues among the top 4 eigenvalues. However, at 18 dB SNR we see nearly 100% improvement of (4,4) MIMO over BF, while (2,4) exhibits about 60% percent improvement over this optimally apodized BF system. We stress that the superior (1,4) curve is the one used in the baseline for the two MIMO gain curves. The bottommost (1,4) curve is for a single omni transmit antenna. The meager performance of this single omni transmit antenna (1,4) system relative to the (1,4) BF is apparent from the capacity curves.

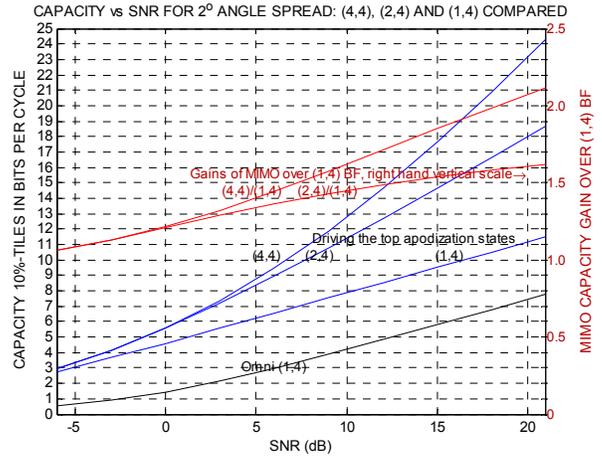

**Figure 12**. Capacity 1%-tiles as a function of SNR for (4,4) and (2,4) MIMO and (1,4) BF versus SNR shown for 2° angular spread and an aperture of one decorrelation length. Progressively greater gains of MIMO over BF with increasing SNR are apparent.

### D. A Closer Look at 6° and 8° Angular Spreads

Figure 13 displays downlink capacity versus SNR curves for a 10%-tile allowed outage. We will see here an example where (4,4) MIMO shows substantial improvement over BF. These curves are for a three decorrelation length aperture with a maximum of 6 array elements per decorrelation length, as, for example would be the case for 6° angle spread in a 1.68m aperture, with a 2 GHz carrier. Alternately, 6° angle spread at a 5 GHz carrier shows superior performance over the 7.5 decorrelation lengths for the same 1.68m.

The other four curves are paired, two blue and two green. The lower curve of each pair is for 2 GHz and the upper is for 5 GHz. The two upper green curves are for (4,4) systems and the lower blue pair are for (1,4) BF which uses the principal APOD for transmission. For both the (1,4) and (4,4) pairs, the 5 GHz system fares better because of the factor of 2.5 decorrelation increase. We see again the over 40% improvement at 2 GHz. As SNRs increase, the MIMO over BF advantage grows and more markedly so at 5 GHz than at 2 GHz. The bottommost black curve is for a (1,4) system with an omni transmit antenna, so that there is no focusing of the downlink signal at the transmitter as there is in the other four curves in the figure.

For (4,4) MIMO, the four top APOD eigenvalues are [10.8302 8.2458 5.7858 4.0082] at 5 GHz, while for 2 GHz they are [8.4010 3.9033 1.8570 1.0204]. The SNR penalties for converting to four equal strength APODs as mentioned in Section VIII.D are just 0.60 dB for 5 GHz and a more substantial 2.54 dB for 2 GHz. This illustrates the general trend of eigenvalues becoming more comparable at greater decorrelation.

As noted in Section VI, 8° angular spread could be more representative for the class of systems of interest in macrocellular applications. Figure 14 displays capacity results for this case, expressing the advantage of (4,4) MIMO over (1,4) BF as a function of aperture size as measured in number of decorrelations lengths. As in the Figure 12, Figure 14 assumes the maximum number of elemental antennas per unit length along the physical aperture as allowed for a 2 GHz carrier, which for a 2° angle spread was 18 antennas per decorrelation length.

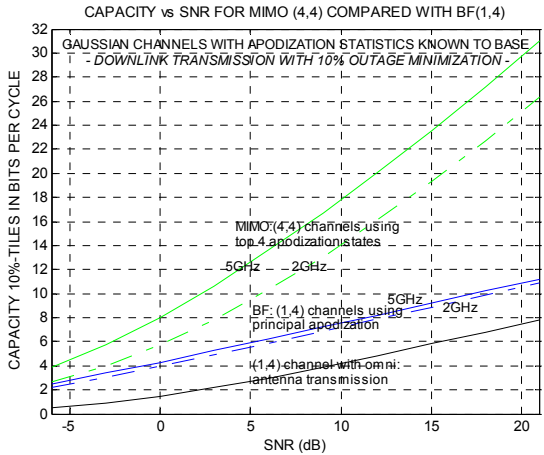

**Figure 13**. For 6° spread, 10%-tiles for (4,4) MIMO (upper pair) over (1,4) BF (next lower pair) are illustrated for a range of SNRs. The lowest curve is the performance of a (1,4) omni-antenna as opposed to BF.

So for an 8° spread we have 18/4 = 4.5 antennas for each decorrelation length. In Figure 14, the horizontal scale starts at 8/9 of a decorrelation length, because that is the minimum length at which 4 elemental antennas fit along the linear aperture, allowing us to speak of a 4 transmit antenna downlink system. Notice that the MIMO spectral efficiency saturation level in Figure 14 is considerably lessened compared to that observed in Figure 12. This is because of the lesser number of antennas per decorrelation length, owing to our ½·λ minimum antenna separation requirement, leaving us with much less antenna gain.

From the 8° MIMO(4,4)/BF(1,4) gain curve shown in Figure 14, we see that compared to the comparable gain curve for the 2° spread case in Figure 12, the gain for 8° is smaller, slightly so at the high end, but markedly so at the low end, where for 2°, at a unit decorrelation length, a 21% improvement was had, whereas, for a 8° spread, Figure 14 shows negligible improvement. The absolute capacities for 8° are also significantly smaller over the decorrelation range than for 2° due to the aforementioned lessened antenna gain, ie, due to the decrease by a factor of 4 in the number of elemental antennas as compared to what we had for 2°. However, it is important to note that the *physical aperture length* at a unit decorrelation for 8° is 4 times less than it is for 2°. For the 1.68 m length at which a 6° spread exhibited a 21% (4,4) MIMO over BF gain, a 8° spread has about a 20% gain. At 3.36m the MIMO (4,4) improvement over BF (1,4) is about 82%.

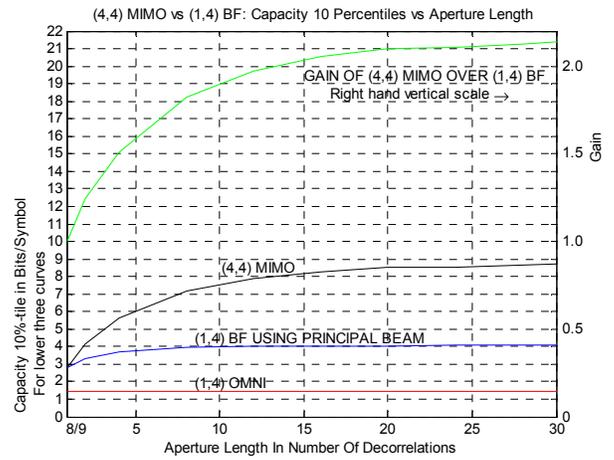

**Figure 14**. For 8° angular spread, the advantage of (4,4) MIMO over (1,4) BF is shown as a function of aperture size as measured in number of decorrelation lengths for 8° spread. These curves are for a 0 dB SNR.

Figure 15 gives capacity curves for 8° spread along the lines of the results presented in Figures 12 and 13. For this important 8° case, three decorrelation lengths are depicted: 2, 4 and 8 corresponding to 0.84 m, 1.68 m and 3.36 m for a 2 GHz carrier. For each of the three curve triplets shown the top curve is for 3.36 m, the middle curve for 1.68 m and the bottom curve of the triplet is for 0.84 m. The red triplet is for (4,4) MIMO and the blue for (1,4) BF while the green curves using the right hand scale show the considerable gain for (4,4) MIMO over (1,4) BF.

The gain grows markedly as SNR increases. At 0dB the gains are already 21%, 51% and 82% for 2, 4, and 8 decorrelation lengths respectively. Observe how the red system of (4,4) MIMO curves rise quickly at the higher SNRs. Indeed, the four degrees of freedom start to be evidenced by the near additional 4 bits/Hz increase at the high SNR end for each 3 dB increase in SNR.

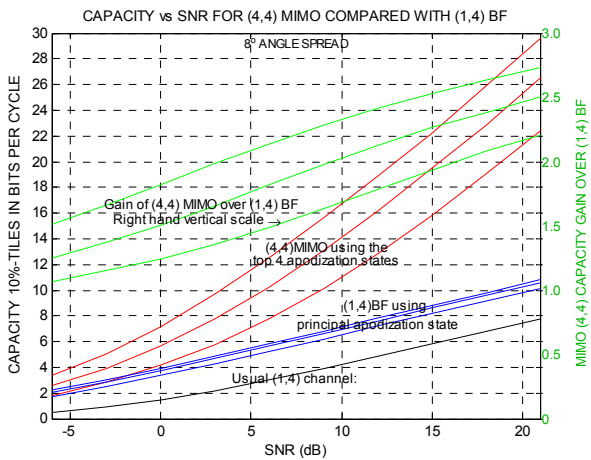

**Figure 15**. Contrast of (4,4) MIMO capacities and (1,4) BF capacities over a wide SNR range for 8° angular spread. The gain of (4,4) MIMO over (1,4) BF is depicted using the scale on the right. Results are given for three aperture lengths: 8, 4 and 2 corresponding to the top, middle and bottom curves respectively.

## X. ALTERNATIVE ARRANGEMENTS

In the arrays considered so far, total flexibility was assumed in signal choice so as to maximize rate. Such flexibility in practical systems incurs additional cost, in the form of a separate (transmit) RF chain per antenna element.

As discussed, high (link) rates may be achieved through the use of statistically independent MIMO transmissions applied to strongest eigenmodes available for a given antenna element arrangement. The strength of each mode scales as the number of antenna elements packed within a coherence length of the channel. MIMO gain over BF scales as the number of "comparable" strength eigenvalues available. Note that when driving only $M$ strongest eigenmodes that are "comparable" the related K-L decomposition of the channel response (17) is then nearly spatially "white". This case arises for large apertures as a consequence of Theorem 2. Whitness enables an arbitrary choice of spatial basis. For the limited angle spread channels considered here, the strongest modes are also lowest order modes, ie, the useful signal subspace is spatially "low-pass". For off-broadside terminals, the modes were spatially modulated by an appropriate (beam steering) phase ramp, resulting in signals that are spatially "bandpass." For this subspace of $M$ strongest eigenmodes, any unitary spatial transformation offers equivalent performance.

One simple alternative arrangement is that of each mode being driven by a spatially disjoint subaperture whose size is on the order of the channel coherence scale. Each such "superelement" offers most of the antenna gain that may be accomplished through BF. For example, it may be observed in Figure 5 that an aperture of 2 decorrelation lengths is 13 dB, ie, within 2 dB of the maximum gain achievable with apertures tending to infinity. Disjoint apertures of such size also enjoy the decorrelation necessary for effective MIMO communication. Decorrelation may be enhanced through introducing gaps between "superelements", as contrasted to the uniform spacing used in preceding sections. A set of $M$ dish antennas, aimed at the user, is one implementation that requires $M$ (say 4) active transmit chains, as opposed to dozens discussed above.

This may be generalized to the case of $L$ users, each at a different azimuth. Here each subaperture may be defined as a passive set of $L$ $\lambda/2$-separated elemental antennas fed through a Butler matrix, which is a hardware implementation of a spatial DFT transform. The number of active (transmit) RF chains is then $LM$, serving $L$ users, each with directional gain of $L$ and a spatial multiplexing gain of $M$. In a multi-user setting and in a system where the total transmit power is fixed, the use of additional antennas allows an $L$-fold increase in system re-use efficiency (at the same SINR per user) and an $M$-fold increase in link multiplexing gain (with $M$-fold reduction in per link SNR).

In particular, for the representative urban environment, characterized by 8° angle spread, a directive antenna array of 1.26 m width (~3 decorrelation lengths at 2 GHz) would achieve most of the directional gain available, as observed from the (1,4) BF capacity curve in Figure 14. Two of such antenna arrays, spaced well apart from each other, would serve as an effective MIMO array, while still being constrained to within 4 m total aperture available at the base station. It may also be pointed out that MIMO gains at higher SNRs become far more pronounced, as illustrated in Figure 12, but for 2° angle spread.

Similar assessment may be made considering using an array of 17 dBi commercial sector antennas (1.5 m high x 0.21m wide) as individual elements. We can compare placing 4 such panel antennas together in beamforming as opposed to using 2 groups of two such antennas in MIMO, with 4-receiver mobiles in both cases. It may be shown that in this case MIMO gives a 20% advantage over BF. Doubling the number of panel antennas to 8 (8 panels for BF vs. 2 groups of 4 panels for MIMO), gives a 50% advantage to MIMO. Total base antenna horizontal apertures are 0.84 m (4 panels) or 1.68 m (8 panels).

## XI. CONCLUSION AND DISCUSSION

We focused on a single communication link. Drawing on propagation basics, we derived a statistical model for the random field along a linear base array associated with a user immersed in clutter. The angular spread of the user was a key parameter. The model served to set the stage for exploring the role of BF as opposed to MIMO, in a context where close, $\lambda/2$ antenna spacing facilitates the formation of beam(s) useful to both BF (one beam) and MIMO (plurality of beams). Basic scaling aspects with respect to angular spread parameter, decorrelation distance and carrier frequency were discussed.

Using K-L expansions which serve in expressing the dominant APODs, capacity at 10% outage was employed as a performance measure. The eigenstates possessed an interesting sine/cosine form which led to insight into how to accommodate the tension between field coherence along the aperture or lack thereof. Asymptotic features of corresponding eigenvalues were discussed because they give insight into how the BF vs MIMO tension will resolve. For MIMO, APOD spectra were explored as a function of the number of transmit antennae and antenna decorrelations along the aperture. Asymptotic features with respect to antenna count and antenna decorrelations were illustrated and explained.

Outage optimization was required, and, for that, *enveloping* was reviewed. A MISO example was used to simply illustrate enveloping and a MISO low outage asymptotic result was uncovered which pointed to extension to the general MIMO case. Using a Monte Carlo enveloping procedure, we compared (1,4) BF to (4,4) MIMO and quantified the emergence of a strong preference for MIMO for larger apertures/shorter decorrelation lengths and larger angular spreads in a single link context. Results were presented for a narrow aperture of $2^o$ where it was difficult to observe when MIMO is favored as compared to BF and for the higher angular spreads of interest $6^o$ and $8^o$, where opportunities for a MIMO preference over BF easily arise. The $8^o$ case is understood be the most relevant case for macrocellular networks.

While this paper provides a basis for future multiuser macrocellular network studies of the BF versus MIMO issue, such a study was beyond the scope of this report. So the numerical examples that we presented were for a single link: For a multiuser network advantages of MIMO over BF are expected to be much more challenging to obtain due to the presence of interference.

For a single link we observed that an aperture length of 3 decorrelations, MIMO (4,4) was better than BF (1,4) by a factor of about 1.7. For a 2 GHz carrier and $2^o$ angular spread, 3 decorrelations could associate with an application using a physically large 5 m aperture or even with a 1.68 m, aperture but at a carrier frequency of 5 GHz along with transmission to a user with $2.4^o$ angular spread.

We saw that the physical aperture length at unit decorrelation for $8^o$ angle spread is 4 times less than it is it is for $2^o$. For a 2 GHz carrier, for the 1.68m length at which a $2^o$ spread had a 21 % (4,4) MIMO over BF gain and an $8^o$ spread has about a 50% gain, while at 3.36 m the MIMO (4,4) improvement over BF (1,4) is about 82%.

Elemental antennas at $\lambda/2$ spacing, spanning a correlated field segment along an aperture partially accounts for the large MIMO and BF capacities that were seen. We stress, that, the 4 transmit antennas, as for the high gain antennas in (4,4) MIMO, could have drawn on 18 or 36 elemental omni antennae if such were needed to constitute the 4 transmit antennas along an aperture 4 or 8 decorrelations long, respectively. Directional gains from such elemental antennas may be thought to be too ambitious an approach, and better antenna designs may be considered.

## XII. ACKNOWLEDGEMENT

We thank our colleagues Antonia Tulino, Stephen Wilkus and Chris Ng for their advice.

## APPENDIX

*Proof of Theorem 3*. DISTRIBUTING EQUAL APODIZATION POWERS MAXIMIZE RATE IN THE ASYMPTOTICALLY SMALL OUTAGE ASYMPTOTE

Here $h_1, h_2, \ldots h_K$ are scalars representing absolute value squared of *iid* complex Gaussian channel power transfer characteristics. Normalizing these k variates by dividing their respective second moments, $v_k$, we obtain:

$$E_k = h_k / v_k, \quad k = 1, 2, \ldots K \quad (A.1)$$

which are i.i.d exponential random variables with density nonvanishing only on $[0,\infty)$ where it is $e^{-t}$. Next, we need $\lambda_1, \lambda_2, \ldots \lambda_K$: which are nonnegative scalars in $[0,P]$ which sum to $P > 0$, these are the per APOD driving powers. In the MISO systems in the text, where the channel is unknown to the transmitter, the argument, x, which appears in the capacity function, LogDet(1 + x) is the random sum

$$x = a_1 E_1 + a_2 E_2 + \ldots a_K E_K \quad (A.2)$$

where $a_k = v_k \lambda_k$, a product of the kth mode driving power $\lambda_k$ and natural mode strength $v_k$.

We will show that in the small outage asymptote it is optimum to put equal power on all K of the APODs that one is allowing to be used. To do so, we recall some facts about adding random variables with independent probability densities and the Laplace transforms of these densities. Then we invoke an induction argument on the number of summands, and, finally, we make some simple optimization observations.

By the well known facts that the Laplace transform of the density of a sum of independent random variables is the product of the transforms of the densities and $a_k E_k$ has the Laplace transform $1/(1+ a_k s)$, we get that the transform of the sum is

$$1/\{(1+ a_1 s)\ (1+ a_2 s)\ \ldots (1+ a_K s)\}. \quad (A.3)$$

Then the inverse transform is a sum of exponentials which decays at infinity as slow as the smallest exponent. Another property that is made clear below is that the pdf at $0^+$ is of the order $t^{K-1}$. That implies that the PDF at $0^+$ is of the order of $t^K$.

We will be interested in the lead term of the power series of the density of the sum in the *neighborhood of* $0^+$, which we abbreviate $N(0^+)$. Clearly for $K = 1$, the lead term is $1/a_1$, and taking the inverse transform for the case $K= 2$ we see that it is $t/(a_1 a_2)$. An induction argument on $K$ shows that for arbitrary $K$ the lead term is given by the expression $t^{K-1}/((a_1 a_2 \ldots a_K) \cdot (K-1)!)$. The logic of the induction is simply seen from the fact that multiplication of transforms corresponds to convolution of densities. By the induction hypothesis, we have the result holding for k in $N(0^+)$. The $k+1$ term, by presence of the $t - \tau$ argument in the convolution presents, in $N(0^+)$, $(1/a_{k+1})\exp(-(\tau-t)/a_{k+1})$. So, in $N(0^+)$ we are estimating

$$\int_0^\tau \frac{1}{a_{k+1} \cdot (k-1)!} \cdot \frac{1}{(a_1 a_2 \ldots a_k)} \cdot t^{k-1} \cdot (1 + \frac{(t-\tau)}{a_{k+1}} + \ldots) dt$$
(A.4)

which clearly yields the desired pdf asymptote.

Since the consequent asymptotic form of the cumulative distribution function in $N(0^+)$ is given by the expression

$$\tau^{K+1} / ((K+1)! \cdot a_1 a_2 \ldots a_{K+1}). \quad (A.5)$$

the next step is to maximize $a_1 a_2 \ldots a_K$, or equivalently to maximize

$$\lambda_1, \lambda_2, \ldots \lambda_K \quad (A.6)$$

subject to the normalized constraint of 1 on the sum of the $\lambda_k$. A simple Lagrange multiplier argument gives that all the $\lambda_k$ should be equal.

Could we have done better to leave out some of the addends of A.2 by allocating them 0 power? We could not, because the more factors present, the higher the power of t at the origin. The higher the power of t the more the first term in the power series hugs the real axis in $N(0^+)$, whatever the positive values of the $v_k$. Hence, the lower the outage in $0^+$. ∎

The above may be extended to the more general case of a sum of K independent variates, where the behavior of each constituent density at the origin is assumed to be well described by a power series:

$$f_k(p_k) = \frac{f_k^{(l_k)} p_k^{l_k}}{l_k!} + ..., \quad (A7)$$

where only the leading term is stated explicitly. Above $l_k$ is the lowest order power of the kth density distribution. Normalizing each non-negative variate $p_k$ by its average power $a_k$, the following quantities may be defined:

$$E_k = p_k/a_k, \quad g(E_k) = a_k f_k(E_k/a_k) \quad (A8)$$

Where $E_k$, as in (A.2), have unit average power. The density (A7) may then be expressed as:

$$f_k(p_k) = \frac{g_k^{(l_k)} p_k^{l_k}}{a_k^{1+l_k} l_k!} + ... \quad (A9)$$

The distribution of the sum of the first 2 variates is given by the convolution, that may be evaluated in closed form [31] for values of $x$ near the origin:

$$\int_0^x dp \frac{g_1^{(l_1)} p^{l_1}}{a_1^{1+l_1} l_1!} \frac{g_2^{(l_2)} (x-p)^{l_2}}{a_2^{1+l_2} l_2!}$$
$$= \frac{g_1^{(l_1)} g_2^{(l_2)} x^{l_1+l_2+1}}{a_1^{1+l_1} a_2^{1+l_2} l_1! l_2!} B(l_1+1, l_2+1) \quad (A10)$$

Where the beta function $B$ may be expressed in terms of ratios of Gamma functions, but that turns out not to have any bearing on power optimization. The convolutions may be continued by induction. The final result for all K terms is:

$$\int_0^x dp \frac{A_{K-1} p^{\sum_{k=1}^{K-2} l_k + K-2}}{a_1^{1+l_1}} \frac{g_2^{(l_K)} (x-p)^{l_K}}{a_K^{1+l_K} l_K!}$$
$$= \frac{A_K g_2^{(l_2)} x^{\sum_{k=1}^{K-1} l_k + K-1}}{\prod_{k=1}^K a_k^{1+l_k}} \quad (A11)$$

The constant $A_K$ is an a product of various derivatives and factorials accumulated with convolution of $K$ terms. For low outage, the cumulative distribution corresponding to (A11) is

$$\frac{A_K g_2^{(l_2)} p^{\sum_{k=1}^{K-1} l_k + K}}{\left(\sum_{k=1}^{K-1} l_k + K\right) \prod_{k=1}^K a_k^{1+l_k}} \quad (A12)$$

Where $a_k = v_k \lambda_k$, as before. Outage probability (A12) is minimized by maximizing $\prod_{k=1}^K a_k^{1+l_k}$, accomplished by selecting mode driving powers $\lambda_k$, subject to constraint $\sum_{k=1}^K \lambda_k = 1$. A Lagrange multipler procedure leads directly to relative mode powers:

$$\lambda_k = \frac{1+l_k}{K + \sum_{k=1}^K l_k} \quad (A13)$$

Where $l_k$ is the lowest order exponent of mode $k$. It may be observed that if all $l_k=0$ (as in the case of exponentially distributed variates), (A13) leads to equal power distribution, as before. Also, modes whose lowest order exponent is higher than that of others would be allocated higher driving power, as a higher exponent reveals a predisposition to higher power realizations.